\documentclass[11pt]{article}
\usepackage[utf8]{inputenc}
\usepackage[T1]{fontenc}
\usepackage[english]{babel}
\usepackage{amssymb,amsmath,fullpage}
\usepackage{graphics}
\usepackage{color}
\usepackage{hyperref}

\usepackage{pstricks,float,fancybox,amssymb,amsmath,graphicx,t1enc,epsfig,psfrag
}
\usepackage{stmaryrd}
\usepackage{amsfonts}
\usepackage{subfigure}
\usepackage{fancyhdr}
\usepackage{fullpage}
\usepackage{float}
\usepackage{latexsym}
\usepackage{placeins}
\usepackage{theorem}
\usepackage[dvips]{geometry}
\usepackage{epsfig}
\usepackage{calc}

\theorembodyfont{\itshape}

\newcommand{\proofendsign}{$\Box$} % \rule{2mm}{2mm}

\newtheorem{thm}{Theorem}[section]
\newtheorem{defi}[thm]{Definition}
\newtheorem{cor}[thm]{Corollary}
\newtheorem{lem}[thm]{Lemma}

\newtheorem{rem}[thm]{Remark}

\newtheorem{qu}{Question}

\newenvironment{pr}{{\noindent \bf Proof
}}{{\hspace*{\fill}\proofendsign\par\bigskip}}
\author{
Rémy Belmonte, Martin Vatshelle \\[.2cm]
University of Bergen, Norway\\
remy.belmonte@ii.uib.no, vatshelle@ii.uib.no
}

\title{ On classes of graphs with logarithmic boolean-width }

\date{}

\begin{document}

%\nocite{*}

\maketitle

\begin{abstract}
Boolean-width is a recently introduced graph parameter.
Many problems are fixed parameter tractable when parametrized by boolean-width,
for instance "Minimum Weighted Dominating Set" (MWDS) problem can be solved in
$O^*(2^{3k})$ time given a boolean-decomposition of width $k$, hence for all
graph classes where a boolean-decomposition of width $O(\log n)$ can be found in
polynomial time, MWDS can be solved in polynomial time.
We study graph classes having boolean-width $O(\log n)$ and problems solvable
in $O^*(2^{O(k)})$, combining these two results to design polynomial algorithms.
We show that for trapezoid graphs, circular permutation graphs, convex graphs,
Dilworth-$k$ graphs, circular arc graphs and complements of $k$-degenerate
graphs, boolean-decompositions of width $O(\log n)$ can be found in polynomial
time.
We also show that circular $k$-trapezoid graphs have boolean-width $O(\log n)$,
and find such a decomposition if a circular $k$-trapezoid intersection model
is given.
For many of the graph classes we also prove that they contain graphs of
boolean-width $\Theta(\log n)$.

Further we apply the results from \cite{boolw2} to give a new polynomial time
algorithm solving all vertex partitioning problems introduced by Proskurowski
and Telle \cite{TP97}. 
This extends previous results by Kratochv\'{i}l, Manuel and Miller \cite{KMM95}
showing that a large subset of the vertex partitioning problems are polynomial
solvable on interval graphs.
\end{abstract}

\section{Introduction}
One of the most studied problems in computer science is the classification of
problems into complexity classes.
For decades there has been done extensive work in order to decide which problems
are solvable in polynomial time (in P).
It is a common belief that NP-complete problems are not in P.
One way to deal with problems we are unable to place in P is to classify on
which inputs they can be solved in polynomial time.
In particular we will in this paper study problems with simple, undirected
graphs as input.
To classify a problem as polynomial on a graph class one have to design an
algorithm that given a graph, in polynomial time either confirms that this graph
is not in the desired class or returns an optimal solution.

One way to generalize the concept of graph classes is by parametrizing the
input.
This means to partition all inputs into classes by assigning them a parameter.
Tree-width, branch-width, clique-width and rank-width are some of the well known
parameters for undirected graphs.
One can then try to design an algorithm that runs in polynomial time for all
graphs with parameter value below some limit $k$.
If we find an algorithm with running time $n^{f(k)}$ we get polynomial
running time for every fixed $k$.
If we find an algorithm with running time $f(k) \times poly(n)$ we get
polynomial
running time even for some graphs with unbounded $k$, this is called an
FPT-algorithm.
In particular if we have an FPT algorithm with running time $2^{O(k)} \times
poly(n)$ we will get polynomial algorithms when $k$ is $O(\log n)$.
We define a triple (graph class $C$, parameter $W$, problem $P$) to be
polynomially parameter tractable (PPT) if:
\begin{enumerate}
 \item Given an $n$-vertex graph $G$, we can in polynomial time compute a
decomposition of $G$ having $W$-width $f(n)$, or conclude that $G$ is not in
the class $C$.
 \item Given a graph $G$ and a decomposition of $G$ having $W$-width $f(n)$ we
can solve problem $P$ on $G$ in polynomial time.
\end{enumerate}

For instance we show that the triple (convex graphs, boolean-width, MWDS) is
PPT.
Note that when both 1) and 2) both can be satisfied with same function $f$ we
can conclude that problem $P$ is polynomial on $C$.
However there are not many well known examples of such PPT triples satisfying
both 1) and 2). We present many such triplets.

Boolean-width is a parameter recently introduced by Bui-Xuan, Telle and
Vatshelle \cite{boolw1}.
In this paper we study the class of graphs with boolean-width $O(\log n)$.
We show that a large class of graphs including interval graphs,
permutation graphs, convex graphs, circular k-trapezoid graphs, Dilworth $k$
graphs and complement of planar graphs have boolean-width $O(\log n)$.
Finally we show how to construct graphs by combining any of these types into
bigger graphs all having boolean-width $O(\log n)$.
For most of these classes we are able to show that the logarithmic bound is
tight up to a constant factor, in particular we show that they contain graphs
having rank-width $\Omega(\sqrt{n})$ and hence boolean-width at least
$\frac{\log n}{2} -O(1)$.
We do not have any polynomial recognition algorithm for graphs of boolean-width
$O(\log n)$ in general.
All of our proofs are constructive, but normally depend on having a certain
representation of the graph as input.
For many of the graph classes discussed in this paper the required
representation can be found in polynomial time meaning we can in polynomial time
build a decomposition given a graph belonging to the graph class, however
some of the required representations are not known to be polynomially
computable.

\begin{figure}[H]\label{fig_classes}
\centering
\includegraphics[scale=.7]{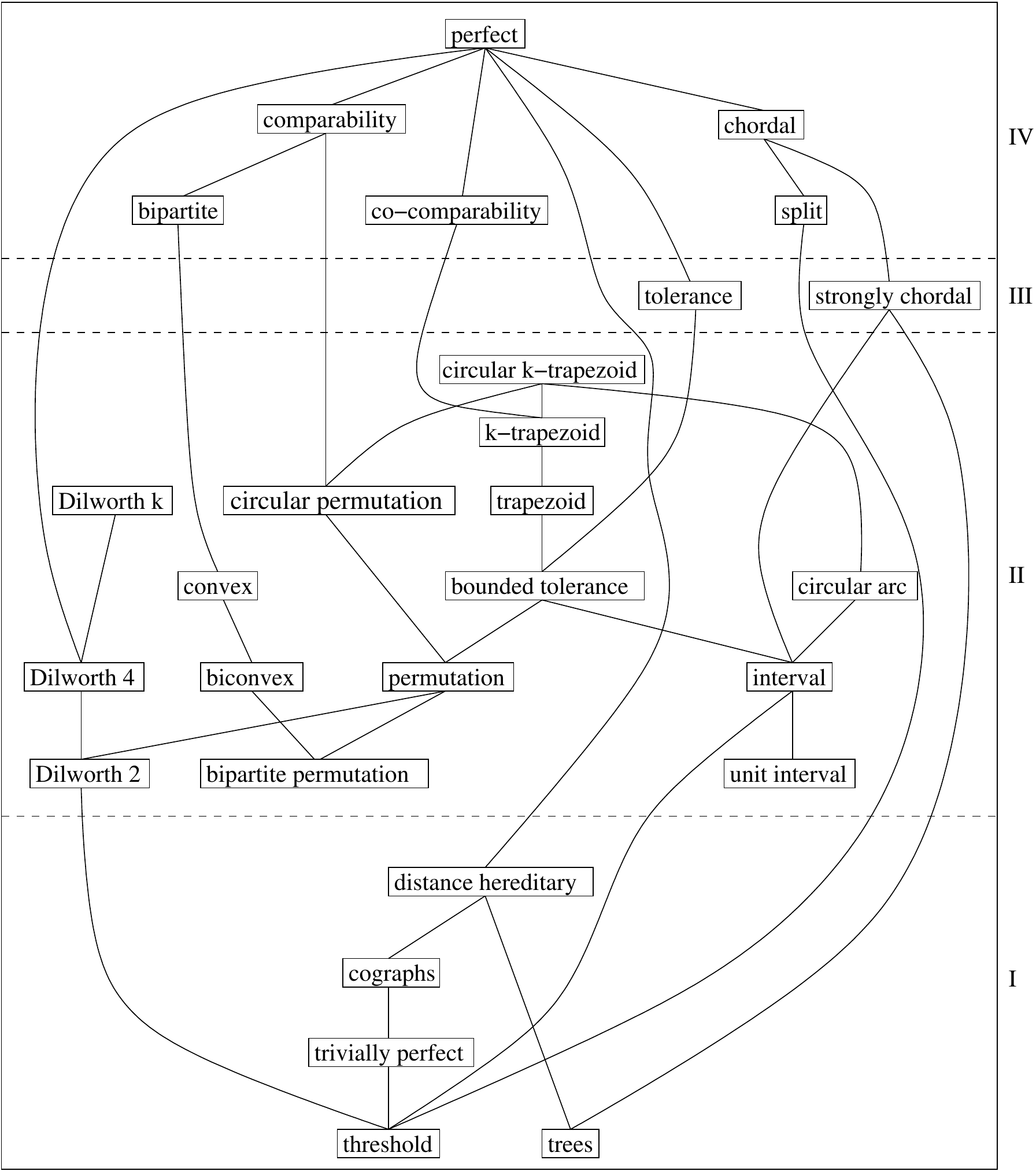}
\caption{
Inclusion diagram of some well-known graph classes. A link between a higher
class A and a lower class B means that B is a subclass of A.
(I) Graph classes where boolean-width is bounded by a constant.
(II) Graph classes having boolean-width $O(\log{n})$.
(III) Boolean-width still unknown.
(IV) There does not exist a boolean-decomposition of value $O(\log{n})$, or it
is NP-complete to compute it.
}
\end{figure}

Combined with the results in \cite{boolw1} this leads to a polynomial algorithm
for weighted dominating set for all the above mentioned graph classes (see
Figure \ref{fig_classes} for an overview).
This unifies algorithms for minimum weighted dominating set on many graph
classes.
In fact we do not know any graph class where weighted dominating set is
polynomial and the boolean-width is not $O(\log n)$.
A constant approximation algorithm for finding a boolean-decomposition of width
$O(\log n)$, would also unify the step of finding boolean decompositions and
hence giving polynomial algorithms also for the classes where representations
can not be found.

The $(\sigma,\rho)$ and vertex partitioning problems which are covered by the
framework
introduced by Proskurowski and Telle \cite{TP97} include among other Independent
Set, Dominating Set, Perfect Code, $k$-Colouring, $H$-Cover and
$H$-Homomorphism.
Bui-Xuan et al.\cite{boolw2} showed that all these problems can be solved in
$2^{O(k^2)} \times poly(n)$ given a boolean-decomposition of width $k$.
Combining this with the results in this paper we get an $O(n^{O(\log n)})$
algorithm.
In order solve these vertex partitioning problems in polynomial time for the
graph classes we have discussed in this paper we must refine the running time
analysis.
In particular one needs to bound the number of $d$-neighbourhoods. 
This is done by bounding the size of a minimal representative.
In all the graph classes we have studied in this paper we were able to build
decompositions such that the minimal representatives have constant size.
Hence all vertex partitioning problems are solvable in polynomial time on
Dilworth $k$ graphs, convex graphs, trapezoid graphs , circular permutation
graphs, circular arc graphs and complement of $k$-degenerate graphs, and also
when given a $k$-trapezoid model or a circular $k$-trapezoid model.

\section{Framework}

When applying divide-and-conquer to a graph we first need to divide the graph.
A common way to store this information is to use a decomposition tree and to
evaluate decomposition trees using a cut function.
The following formalism is refered to as branch decomposition of a cut function
and is standard in graph and matroid theory (see, e.g.,~\cite{OS06,GGW02,GM10}).
Throughout the paper we will for $A \subseteq V(G)$ let $\overline{A}$ denote
the set $ V(G) \setminus A$.

\begin{defi}\label{def_f_width}
\emph{A decomposition tree of a graph $G$ is a pair $(T,\delta)$ where $T$ is a
tree having internal nodes of degree three and $n=|V(G)|$ leaves, and $\delta$
is a bijection between the vertices of $G$ and the leaves of $T$.
Every edge of $T$ defines a cut $\{A,\overline{A}\}$ of the graph, i.e. a
partition of $V(G)$ in two parts, namely the two parts given, via $\delta$, by
the leaves of the two subtrees of $T$ we get by removing the edge.
Let $f:2^V\rightarrow\mathbb{R}$ be a symmetric function, i.e.
$f(A)=f(\overline{A})$ for all $A\subseteq V(G)$, also called a \emph{cut
function}.
The $f$-width of $(T,\delta)$ is the maximum value of $f(A)$, taken over all
cuts $\{A, \overline A\}$ of $G$ given by an edge $uv$ of $T$.
The $f$-width of $G$ is the minimum $f$-width over all decomposition trees of
$G$.}
\end{defi}

Caterpillar decompositions are decompositions were the underlying tree is a path
with one leaf added to every internal node of the path.
Many of our proofs will construct caterpillar decompositions. To describe a
caterpillar decomposition, we only give an ordering of the vertices. 
To construct the caterpillar decomposition from an ordering, follow the path of
the caterpillar and map the vertices to the leafs attached to the path.

The cuts $\{A,\overline{A}\}$ given by edges of the decomposition tree are used
in the divide step of a divide-and-conquer approach.
We solve the problem recursively, following the edges of the tree $T$ (after
choosing a root) in a bottom-up fashion, on the graphs induced by vertices of
one side and of the other side of the cuts. 
In the conquer step we must join solutions from the two sides, and this is
usually the most costly and complicated operation.
Normally the conquer step uses Dynamic programming.
The question of what 'solutions' we should store to get an efficient conquer
step is related to what type of problem we are solving.

We suggest that the following equivalence relation on subsets of $A$ will be
useful for solving problems like dominating set.

\begin{defi}\label{neighbourEQ}
\emph{Let $G$ be a graph and $A\subseteq V(G)$.
Two vertex subsets $X \subseteq A$ and $X' \subseteq A$ are {\sl neighbourhood
equivalent} w.r.t.\ $A$, denoted by $X \equiv_A X'$, if $\overline A\cap N(X) =
\overline A\cap N(X')$.}
\end{defi}

In order to bound the number of neighbourhood equivalence classes we define boolean-width as follows:

\begin{defi}[Boolean-width]
\label{def:booleanWidth}
The $cut\textrm{-}bool:2^{V(G)}\rightarrow\mathbb{R}$ function of a graph $G$ is
$$cut\textrm{-}bool(A)=\log_2|\{S\subseteq \overline{A}:~\exists X\subseteq A
\wedge ~ S = \overline{A} \cap \bigcup_{x\in X}N(x)\}|$$
Using Definition~\ref{def_f_width} with $f=cut\textrm{-}bool$ we define the
boolean-width of a decomposition tree, denoted $boolw(T,\delta)$, and the
boolean-width of a graph, denoted $boolw(G)$.
\end{defi}

It is known from Boolean matrix theory that $cut\textrm{-}bool$ is symmetric~\cite{K82}.

Note that we take the logarithm base 2 of the number of equivalence classes
simply to ensure that $0 \leq boolw(G)\leq |V(G)|$, which will ease the
comparison of boolean-width to other parameters.

Rank-width was introduced by Oum and Seymour in \cite{thesisOum, OS06}. One way
of defining rank-width is to replace the union in the definition of
boolean-width by symmetric difference. 
The symmetric difference operator $\bigtriangleup$ applies to a family of
sets and returns the set of elements appearing in an odd number of sets.

\begin{defi}[Rank-width]
\label{def:rankWidth}
The $cut\textrm{-}rank:2^{V(G)}\rightarrow\mathbb{N}$ function of a graph $G$ is
$$cut\textrm{-}rank(A)=\log_2|\{S\subseteq \overline{A}:~\exists X\subseteq A
\wedge ~ S =\overline{A} \cap \bigtriangleup_{x\in X}N(x)\}| =
rk(M_{A,\overline{A}})$$
Using Definition~\ref{def_f_width} with $f=cut\textrm{-}rank$ we define the
rank-width of a decomposition tree, denoted $rw(T,\delta)$, and the rank-width
of a graph, denoted $rw(G)$.
\end{defi}

Now we establish some terminology used in this paper.

\begin{defi}
 Let $G = (V,E)$ be a graph, $(A,\overline{A})$ a cut of $G$. Let $S = A \cap
N(\overline{A})$. We define the middle vertices of a cut $m(A)$ as the
vertices found by the following procedure:\\
Let $S' = S$. 
While there are vertices $u,v \in S'$ such that $N(u) \setminus A =
N(v) \setminus A$ remove $u$ from $S'$.
Return $S'$.
\end{defi}

Note that $|m(A)| \leq n - 1$.

Many of the graphs we study in this paper are defined via intersection models.
\begin{defi}
An intersection model of a graph is a one to one mapping of objects to the
vertices of a graph such that there is an edge between two vertices in the graph
if and only if the objects mapped to the vertices intersect. Geometrical
objects and sets are the most common examples of such objects.
\end{defi}

\section{Upper Bounds on Boolean-width of Graph Classes}
\label{upper}
\subsection{Permutation graphs}

\begin{defi}
A graph is a permutation graph if and only if it has an intersection model
consisting of straight lines (one per vertex) between two parallels.
\end{defi}

For more information on permutation graphs see \cite{perm1}.

\begin{thm} \label{thm_perm}
The boolean-width of a permutation graph $G$ is at most $\log n$.
\end{thm}

\begin{pr}
We build a caterpillar decomposition by sorting the vertices by the upper
endpoint of their corresponding line.
Let us now consider a cut $(A,\overline{A})$ of the decomposition.
Let $\sigma$ be the total ordering of the vertices of $m(A)$ sorted by their
lower endpoint, hence $\forall u,v \in m(A), \sigma(u) \leq \sigma(v)$ iff the
lower endpoint of $u$ is to the left of the lower endpoint of $v$.
Since all upper endpoints of lines corresponding to vertices of $A$ are to the
left of all upper endpoints of lines corresponding to vertices of
$\overline{A}$, two vertices $u \in A, u' \in \overline{A}$ are neighbours iff
the lower endpoint of $u$ is to the right of the lower endpoint of $u'$.
Hence for any set $S \subseteq A$ there exists $x \in S$ such that $N(S) \cap
\overline{A} = N(x) \cap \overline{A}$, namely the vertex of $S$ with the
rightmost lower endpoint.
Then there are at most $m(A)$ neighbourhoods, and the theorem holds.
\end{pr}

\begin{lem} \cite{perm3}
Given a graph, one can in linear time either decide that the
graph is not a permutation graph, or output a permutation model for the graph.
\end{lem}

\subsection{Circular permutation graphs}

\begin{defi}
A graph is a circular permutation graph if it has an intersection model
consisting of curves between two distinct concentric circles, such that no two
curves cross in more than one point, and no two curves touch without crossing.
\end{defi}

For more information on circular permutation graphs see \cite{circperm1}.

\begin{thm} \label{thm_circleprem}
The boolean-width of a circular permutation graph $G$ is at most $2 \log n$.
\end{thm}

A bound of $4 \log n$ follows from Theorem \ref{thm_circlektrapez}.
The proof of the improved bound is similar to Theorem \ref{thm_perm}.

\begin{lem} \cite{circperm2}
Given a graph, one can in polynomial time either decide that the
graph is not a circular permutation graph, or output a circular permutation
model for the graph.
\end{lem}

\subsection{k-Trapezoid graphs}

\begin{defi}
A graph is a $k$-trapezoid graph if it is the intersection graph of
$k$-trapezoids, where a $k$-trapezoid is given by $k$ intervals on $k$
parallel lines.
\end{defi}

For more information on $k$-trapezoid graphs see \cite{Flotow95}.

\begin{thm} \label{thm_ktrapez}
The boolean-width of a $k$-trapezoid graph $G$ is at most $k \log n$.
\end{thm}

The proof is similar to Theorem \ref{thm_circlektrapez} and therefore was moved
to the appendix.

It is NP-complete to compute a $k$-trapezoid model for a $k$-trapezoid graph for
$k \geq 3$ (Yannakakis \cite{Yannakakis82} and Flotow \cite{Flotow95}).

Interval graphs are exactly the $1$-trapezoid graphs. 

\begin{cor} \label{cor_interval}
 The boolean-width of an interval graph $G$ is at most $\log C$, where $C$ is
the size of the biggest clique in $G$.
\end{cor}

This holds since any set of intervals that contain a common point forms a
clique. A proof can be found in appendix.

\subsection{Circular k-trapezoid graphs}

\begin{defi}
A graph is a circular $k$-trapezoid graph if it is the intersection graph of
circular $k$-trapezoids, where a circular $k$-trapezoid is given by $k$
intervals on $k$ concentric circles.
\end{defi}

For more information on circular $k$-trapezoid graphs see \cite{S03}.

\begin{thm}\label{thm_circlektrapez}
The boolean-width of a circular $k$-trapezoid graphs graph $G$ is at most
$2k \log n$.
\end{thm}

\begin{pr}
We build a caterpillar decomposition by starting on a point $p$ of the circle
and by adding vertices of which $k$-trapezoid contains the point $p$ of the
innermost circle. 
We then order the $k$-trapezoids not containing $p$ by the distance from $p$ to
the point of the $k$-trapezoid closest to $p$, and add the vertices in that
order.
Now, we look at a cut $(A,\overline{A})$ of the decomposition.
We show that the boolean-width of any circular $k$-trapezoid graph is bounded
by showing that any neighbourhood has a representative of size at most $2k$.
Let us take any $S \subseteq A$ with $|S| > 2k$, we build a set $S' \subseteq S$
with $N(S) \cap \overline{A} = N(S') \cap \overline{A}$ and $|S'| \leq 2k$:
For each line $i$, we take the 2 $k$-trapezoids $r_i$ and $l_i$
such that $r_i$ is the $k$-trapezoid containing the point on
the $i^{th}$ line furthest in clockwise direction and similarly we choose $l_i$
in counter-clockwise direction. 
We set $S'$ as the set of all $r_i$ and $l_i$.
Let us assume for contradiction that $\exists x \in \overline{A}: x \in N(S)
\backslash N(S')$. 
There must exist some $j$ such that the $k$-trapezoid of $x$ intersects some
trapezoid of $S$ on the $j^{th}$ line. 
Either it intersects $r_j$ or $l_j$, else the whole trapezoid of $x$ is
contained in the area between $r_j$ and $l_j$, then by construction of the
decomposition, $x$ would have been in $A$.
Thus $N(S) \cap \overline{A} = N(S') \cap \overline{A}$, and hence any
neighbourhood has a representative of size at most $2k$ and the theorem
holds.
\end{pr}

Circular arc graphs are exactly circular $1$-trapezoid graphs.

\begin{cor} \label{cor_circarc}
 The boolean-width of a circular arc graph $G$ is at most $2 \log C + 2$, where
$C$ is the size of the biggest clique in $G$.
\end{cor}

This holds since any set of arcs that contain a common point forms a
clique. A proof can be found in appendix.

\subsection{Convex graphs}

\begin{defi}
An ordering $<$ of X in a bipartite graph $B=(X,Y,E)$ has the adjacency property
if for every vertex $y$ in $Y$, $N(y)$ consists of vertices that are consecutive
(an interval) in the ordering $<$ of $X$.
A bipartite graph $(X,Y,E)$ is convex if there is an ordering of $X$ or $Y$ that
satisfies the adjacency property.
\end{defi}

For more information about convex graphs see \cite{convex1}.

\begin{thm}
The boolean-width of a convex graph $G$ is at most $\log n$.
\end{thm}

\begin{pr}
Since $G$ is convex we know there exists a bipartition $(X,Y)$ of $V$ and
$\sigma_X$ an ordering of $X$ such that $\forall u \in Y, x,y \in N(u)$ we have
$\forall z \in X : \sigma_X(x) < \sigma_X(z) < \sigma_X(y)$ then $z \in N(u)$.
We construct a total ordering $\sigma$ of $V$ from $\sigma_X$ by keeping the
ordering of vertices in $X$ and for each vertex $v \in Y$ we insert $v$
immediately after the last element of $N(v)$. 
We construct a caterpillar decomposition based on the order $\sigma$.
Consider a cut $(A,\overline{A})$ in the caterpillar decomposition. 
Then $m(A) \cap Y = \emptyset$ by construction of $\sigma$.
 
Let $v_1, v_2, \dots, v_t$ be the ordering of the vertices of $m(A)$ induced by
$\sigma$, ($t = |m(A)|)$.
Since all the vertices in $Y \cap \overline{A}$ appear later in $\sigma$
than $v_t$, they all either see $v_t$ or have no neighbours in $A$. 
By the property of a convex graph if some vertex $u \in Y \cap \overline{A}$ see
$v_i$ then $u$ also see $v_{i+1}$.
Hence we get $boolw(G) \leq \log(|m(A)|+1)$ and since $|m(a)| \leq n-1$ the
theorem holds.
\end{pr}

\begin{lem} \cite{convex2}
Given a graph, one can in polynomial time either decide that the
graph is not a convex graph, or output an ordering verifying that the graph is
convex.
\end{lem}

\subsection{Graphs of bounded Dilworth number}

\begin{defi}[Dilworth number]
Two vertices $x$ and $y$ are said to be comparable if either $N(y) \subseteq
N[x]$ or $N(x) \subseteq N[y]$. The Dilworth number of a graph is the largest
number of pairwise incomparable vertices of the graph. A graph is a Dilworth $k$
graph if it has Dilworth number $k$.
\end{defi}

In order to prove our next result, we will need the following theorem, well
known as Dilworth's theorem from posets theory:

\begin{thm}[Dilworth \cite{dilworth-thm}]
In a finite partial order, the size of a maximum anti-chain is equal to the
minimum number of chains needed to cover its elements.
\end{thm}

\begin{thm}\label{dilworth-bound}
The boolean-width of a Dilworth $k$ graph $G$ is at most $k \log n$.
\end{thm}

\begin{pr}
From Dilworth's theorem, we know that if a graph $G$ is a Dilworth $k$, then the
Hasse diagram associated with the inclusion of neighbourhoods of $G$ can be
covered with $k$ chains. We call these chains $C_1,...,C_k$.
Thus, we build a caterpillar decomposition by adding the vertices of $C_1$,
then the vertices of $C_2$, and so on.
Let us now take any cut $(A,\bar{A})$ of the decomposition.
$\forall S \subseteq A,$ we call $X_S$ the subset of $S$ such that $\forall x
\in S, \exists y \in X_S : N(x) \subseteq N(y)$. Thus we have $N(S) =
N(X_S)$. Moreover, $|X_S| \leq k$ (at most one vertex for each chain).
Therefore there can be at most $n^{k}$ neighbourhoods and the result holds.
\end{pr}

\subsection{Complements of k-degenerate graphs}

\begin{defi}
A graph $G$ is $k$-degenerate if and only if there exists an elimination
ordering on the
vertices such that every vertex has at most $k$ neighbours appearing later in the ordering.
\end{defi}

\begin{thm}
The boolean-width of a graph $G$, where $\overline{G}$ is a $k$-degenerate
graph, is at most $k \log n$.
\end{thm}

\begin{pr}
We build a caterpillar decomposition of $G$ using the elimination ordering
induced by the $k$-degeneracy of $\overline{G}$.
We consider a cut $(A,\overline{A})$ of the decomposition. Since $\overline{G}$
is $k$-degenerate, we have $\forall x \in A, d(x) \leq k$ in $\overline{G}$,
therefore $\forall x \in A, |N(x) \cap \overline{A}| \geq |\overline{A}| -
k$ in
$G$, and thus $|\{ N(x) \cap \overline{A} : X \subseteq A \}| \leq
\binom{n}{k}$. Hence, we get
the result $boolw(G) \leq \log \binom{n}{k} \leq k \log n$.
\end{pr}

\section{Lower Bounds on Boolean-width of Graph Classes}

In the previous section we showed many classes of graphs with boolean-width
$O(\log n)$, now we want to show that a big class of graphs have low
boolean-width and high rank-width.\\
A Hsu-graph is a bipartite graph $H = (V,E)$ where $V = \{v_1,v_2,\dots
v_a\},\{u_1,u_2,\dots u_b\}$ where $v_i,u_j \in E(H) \Leftrightarrow i \leq
j$.\\
A Hsu-join-chain of length $q$ and width $p$ is constructed as follows:
Let $\mathcal{F} = G_1, G_2, \dots, G_q$ be a family of connected graphs, all on
at least $p$ vertices and boolean-width $O(\log p)$. For each graph $G_i$ pick a
set $S_i$ of $p$ vertices. For each pair $G_i,G_{i+1}$ connect $S_i$ to
$S_{i+1}$ by a Hsu-graph.

\begin{thm}
Let $G$ be a HSU-join-chain of length $q$ and width $p$ where $q > 3p$ then
$boolw(G) \in O(\log p)$ and rank-width$(G) \geq p/2$.
\end{thm}

\begin{pr}
Let $\mathcal{F} = G_1, G_2, \dots, G_q$ be the family of graphs used to
construct $G$.
For each graph $G_i$ take an optimal rooted decomposition tree and identify the
roots with leafs of a caterpillar using the order $G_1, G_2, \dots, G_q$.
For a cut $(A,\overline{A})$ where $A \subseteq V(G_i)$ for some $i$ we know
there are $2^{O(\log p)}$ neighbourhoods in the cut $(A, V(G_i) \setminus A)$,
$O(p)$ neighbourhoods in the cut $(A, V(G_{i-1}))$ and $O(p)$ neighbourhoods in
the cut $(A, V(G_{i+1}))$.
This means that boolean-cut value of $(A,\overline{A})$ is $O(\log p)$.
For the cuts $(V(G_1) \cup V(G_2) \cup \dots \cup V(G_i), V(G_{i+1}) \cup \dots
\cup V(G_i) )$ only $V(G_i) \cup V(G_{i+1})$ have neighbours on each side and
hence the cut edges form a Hsu-graph and hence the boolean-width of $G$ is
$O(\log p)$.
To show that $G$ has high rank-width we use the fact that there must be a
$(\frac{1}{3},\frac{2}{3})$-balanced cut $(A,\overline{A})$ and we show that
every such cut has rank-width at least $p/2$.
Assume for contradiction that the rank-width of $(A,\overline{A})$ is less than
$p/2$, then there are at most $p$ graphs in $\mathcal{F}$ intersecting both $A$
and $\overline{A}$.
Since otherwise in each in each $G_i$ intersecting both sides of the cut we can
find a crossing edge.
Taking the edges from all $G_i$ such that $i$ is odd (resp. even) will give an
induced matching of size at least $p/2$ and hence the rank-width of the cut
would be at least $p/2$.
Since the cut is balanced we know that there is some graph in $\mathcal{F}$
completely contained in $A$ and some graph in $\mathcal{F}$ completely contained
in $\overline{A}$. Otherwise rank-width would be at least $q/6 > p/2$.
For each cut $(G_i,G_{i+1})$ we know that there is an induced subgraph of the
cut isomorphic to a Hsu-graph of size $|S_{i+1} \cap A|-|S_i \cap A|$.
Combining all the Hsu-graphs obtained from either the cuts with $i$ odd or $i$
even ensures that the cut has rank-width at least $p/2$.
\end{pr}

\begin{cor} \label{cor_HsuSC}
 There exists an infinite family of bipartite permutation graphs with rank-width
$\Omega(\sqrt{n})$.
\end{cor}

Let a Hsu-Stable-chain of length $q$ and width $p$ be the Hsu-join-chain of
length $q$ and width $p$ where $\forall i, G_i$ is a stable set of size $p$.
A Hsu-Stable-chain is a bipartite permutation graph with boolean-width
$\Theta(n)$.

\begin{figure}[H]
\centering
\subfigure[]{
\includegraphics[scale=.65]{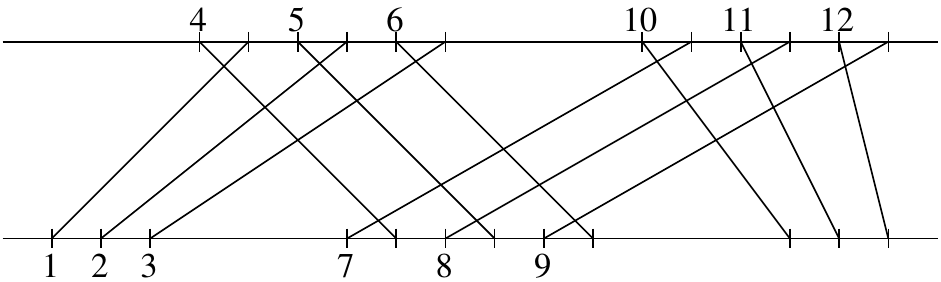}
\label{rank-bipperm}}
\hspace{1cm}
\subfigure[]{
\includegraphics[scale=.65]{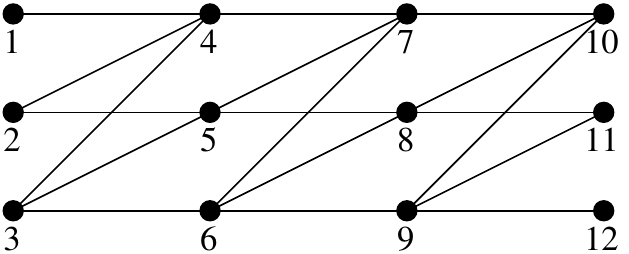}
\label{rank-bipperm-graph}}
\caption{Hsu-Stable chain $3 \times 4$ and its permutation representation.}
\end{figure}

\begin{cor} \label{cor_HsuCC}
 There exists an infinite family of unit interval graphs (i.e.
an interval graph having all intervals of unit length) with rank-width
$\Omega(\sqrt{n})$.
\end{cor}

Let a Hsu-Clique-chain of length $q$ and width $p$ be the Hsu-join-chain of
length $q$ and width $p$ where $\forall i, G_i$ is a Clique of size $p$.
A Hsu-Clique-chain is a unit interval graph with boolean-width $\Theta(n)$.

\begin{figure}[H]
\centering
\subfigure[]{
\includegraphics[scale=0.85]{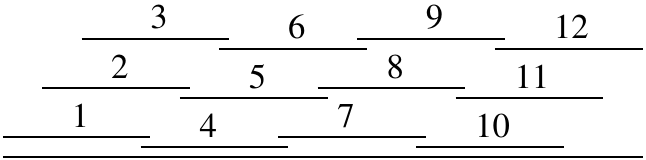}
\label{rank-unitint}}
\hspace{1cm}
\subfigure[]{
\includegraphics[scale=.65]{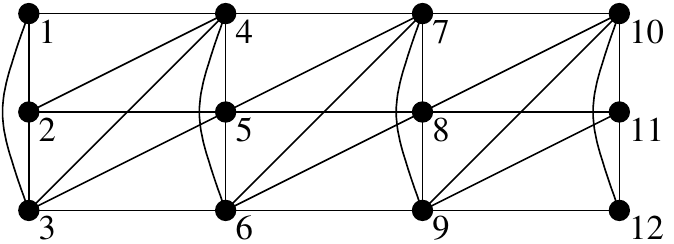}
\label{rank-unitint-graph}}
\caption{Hsu-Clique chain $3 \times 4$ and its unit interval representation.}
\end{figure}

\begin{rem}\label{no-log}
Let $G$ be a graph which belongs to a class where there exists a problem
solvable in time $O^*(2^{O(boolw(G))})$ which is NP-complete, then either $G$
does not have a boolean decomposition of value $O(\log n)$, or it is
NP-complete to build such a decomposition.
\end{rem}
It implies that, unless P=NP, one cannot compute a boolean
decomposition of
value $O(\log n)$ in polynomial time for any of the following graph classes:
\begin{itemize}
 \item Split graphs (from \cite{Bertossi84})
 \item Circle graphs (from \cite{MDScircle})
 \item Co-comparability graphs (from \cite{Chang95})
 \item Chordal bipartite graphs (from \cite{MDSchordbip})
\end{itemize}

\section{Vertex partitioning problems}

In \cite{TP97} Telle and Proskurowski introduced a generalized framework for
handling many types of vertex subset and vertex partitioning problems in a
unified manner. These types of problems have been studied in many ways. Normally
they are described by a degree constraint matrix although there are also
problems not describable by a degree constraint matrix.

\begin{defi}
\label{partitioning}
 A {\sl degree constraint} matrix $D_q$ is a $q$ by $q$ matrix with entries being 
 finite or co-finite subsets of natural numbers.
 A {\sl $D_q$-partition} in a graph $G$ is a partition $\{V_1,V_2,...,V_q\}$ of 
 $V(G)$ such that for $1 \leq i,j \leq q$ we have 
 $\forall v \in V_i: |N(v) \cap V_j| \in D_q[i,j]$.
\end{defi}

We call the vertex partitioning problems describable by a degree constraint
matrix for $D_q$-problems.
Telle and Proskurowski showed that all $D_q$-problems are solvable in FPT time
parametrized by tree-width\cite{TP97}.
Kobler and Rotics showed that $D_q$-problems are solvable on graphs of bounded
clique-width\cite{KR01}, and with a little effort their algorithm can be made
into an FPT algorithm. 
Bui-Xuan et al showed that $D_q$-problems are FPT when parametrized by
boolean-width\cite{boolw2}. 
Kratochvil et al. \cite{KMM95} showed that a large subset of the $D_q$-problems
are solvable in polynomial time on interval graphs.
We generalize the results of \cite{KMM95} by showing that all $D_q$-problems
are solvable in polynomial time on many well known graph classes.

We will build on the algorithm of Bui-Xuan\cite{boolw2}, there the bottleneck
for running time is the number of equivalence classes of $d$-neighbourhoods.

\begin{defi}[$d$-neighbour equivalence]\label{dNeighbour}
\emph{Let $G$ be a graph and $A\subseteq V(G)$ a vertex subset of $G$.
Two vertex subsets $X \subseteq A$ and $X' \subseteq A$ are {\sl $d$-neighbour
equivalent} w.r.t.\ $A$, denoted by $X \equiv^d_A X'$, if\\
$~\hfill\forall v \in \overline{A} : \left(|N(v) \cap X|\frac{}{}=\frac{}{}|N(v)
\cap X'|\right) \vee \left(|N(v) \cap X| \geq d\frac{}{} \wedge\frac{}{} |N(v)
\cap X'| \geq d \right).\hfill~$}

\end{defi}

The integer value $d$ depends on the sets used in the degree constraint matrix.
Let $d(\mathbb{N}) = 0$.
For every finite or co-finite non-empty set $\mu \subseteq \mathbb{N}$, let
$d(\mu) = 1 + \min ( \max x : x \in \mu, \max x : x \notin \mu )$.
For a matrix $D_q$, the value $d$ will be $\max d(\mu) : \mu \in D_q$. 

Let $UN_d$ be the maximum number of equivalence classes of the $d$-neighbourhood
equivalence relation over the cuts of a decomposition, the running time of the
algorithm given in \cite{boolw2} will be $O(n(m + qd * boolw(G) * UN_d^3 *
2^{boolw(G)}))$.
What we need in order to give a polynomial algorithm for all $D_q$-problems on
a specific graph class is to give a decomposition with $UN_d$ bounded by some
polynomial in $n$ for each cut and boolean-width by $O(\log{n})$.
One way to do this is to show that every $d$-neighbourhood has a representative
of constant size in the $O(\log n)$ boolean-decomposition.

In all the proofs of this paper we bound the boolean-width by bounding the size
of the representatives needed.
More formally, for every $A \subseteq V(G)$ given by the decomposition tree, we
showed that for every $S \subseteq A$ there exists a set $R \subseteq S$ such
that $N(S) \cap \overline{A} = N(R) \cap \overline{A}$.
In order to bound boolean-width we do not need $R \subseteq S$, but this will be
crucial to bound the number of $d$-neighbourhoods.

\begin{lem}
Given $A \subseteq V(G)$, assume $\forall S \subseteq A, \exists R \subseteq S$
such that $N(S) \cap \overline{A} = N(R) \cap \overline{A}$ and there exists an
integer $k$ such that $|R| \leq k$. Then, $\exists R' \subseteq S$ such that $R'
\equiv_A^d S$ and $|R'| \leq dk$.
\end{lem}
\begin{pr}
 Proof by induction on $d$. $1$-neighbourhoods are exactly the same as
neighbourhoods, hence the lemma is trivially true for $d=1$.
 Assume the statement of the lemma is true for all values up to $d-1$. 
 Let $S$ and $R$ be sets satisfying the conditions in the lemma.
 Find minimal $X \subseteq S$ such that $S \setminus R \equiv_A^{d-1} X$, then
by induction hypothesis, $|X| \leq (d-1)k$.
 Let $R' = R \cup X$. Now it is easy to see that $R' \equiv_A^d S$.
\end{pr}

Combining the above lemma with the results in Section \ref{upper}, especially
the fact that neighbourhoods can be described by a representative of constant
size we get the following theorem.

\begin{thm}
All $D_q$-problems are solvable in polynomial time on the following graph classes:
Dilworth $k$ graphs, convex graphs, trapezoid graphs , circular
permutation graphs, circular arc graphs and complement of $k$-degenerate graphs.
\end{thm}

\begin{thm}
Given a $k$-trapezoid model or a circular $k$-trapezoid model of $G$ as input.
All $D_q$-problems are solvable in polynomial time on $G$.
\end{thm}

\section{Conclusion}

We have shown that a large family of graph classes including interval graphs and
trapezoid graphs have low boolean-width on one hand, while on the other hand,
they have high rank-width and we can easily find a decomposition of low
boolean-width.
This means for instance that the best algorithms for the minimum dominating set
problem using boolean-decompositions are polynomial while the best known using
rank-decompositions are not.
Moreover, since the size of the representative is bounded by a constant for all
these graph classes, we can improve the analysis of the running-time of the
algorithm for vertex partitioning problems in \cite{boolw2}.
Thus, we have provided many graph classes for which every finite vertex
partitioning problem can be solved in polynomial time, as well as their weighted
version.

\begin{qu}
Is there any graph class having boolean-width $\Omega(\log n)$ where weighted
dominating set is polynomially solvable?
\end{qu}

\begin{qu}
For tolerance graphs, both the complexity class of Dominating Set and the value
of boolean-width are unknown.
\end{qu}

%\begin{scriptsize}
\bibliographystyle{plain}
\bibliography{biblio}

\begin{thebibliography}{10}

\bibitem{boolw2}
Isolde Adler, Binh-Minh Bui-Xuan, Yuri Rabinovich, Gabriel Renault, Jan~Arne
  Telle, and Martin Vatshelle.
\newblock On the boolean-width of graphs : structure and applications.
\newblock In {\em WG}, 2010.

\bibitem{Bertossi84}
Alan~A. Bertossi.
\newblock Dominating sets for split and bipartite graphs.
\newblock {\em Inf. Process. Lett.}, 19(1):37--40, 1984.

\bibitem{convex2}
Kellogg~S. Booth and George~S. Lueker.
\newblock Linear algorithms to recognize interval graphs and test for the
  consecutive ones property.
\newblock In {\em STOC '75: Proceedings of seventh annual ACM symposium on
  Theory of computing}, pages 255--265, 1975.

\bibitem{boolw1}
Binh-Minh Bui-Xuan, Jan~Arne Telle, and Martin Vatshelle.
\newblock Boolean-width of graphs.
\newblock In {\em IWPEC}, pages 61--74, 2009.

\bibitem{Chang95}
Maw-Shang Chang.
\newblock Weighted domination on cocomparability graphs.
\newblock In {\em ISAAC}, pages 122--131, 1995.

\bibitem{trapez1}
Ido Dagan, Martin~C. Golumbic, and Ron~Y. Pinter.
\newblock Trapezoid graphs and their coloring.
\newblock {\em Discrete Applied Mathematics}, 21(1):35--46, 1988.

\bibitem{dilworth-thm}
Robert~P. Dilworth.
\newblock A decomposition theorem for partially ordered sets.
\newblock {\em The Annals of Mathematics}, 51(1):161--166, 1950.

\bibitem{perm1}
Shimon Even, Amir Pnueli, and Abraham Lempel.
\newblock Permutation graphs and transitive graphs.
\newblock {\em J. ACM}, 19(3):400--410, 1972.

\bibitem{Flotow95}
Carsten Flotow.
\newblock On powers of m-trapezoid graphs.
\newblock {\em Discrete Applied Mathematics}, 63(2):187--192, 1995.

\bibitem{GGW02}
James~F. Geelen, Bert Gerards, and Geoff Whittle.
\newblock Branch-width and well-quasi-ordering in matroids and graphs.
\newblock {\em J. Comb. Theory, Ser. B}, 84(2):270--290, 2002.

\bibitem{MDScircle}
J.~Mark Keil.
\newblock The complexity of domination problems in circle graphs.
\newblock {\em Discrete Appl. Math.}, 42(1):51--63, 1993.

\bibitem{K82}
Kane~H. Kim.
\newblock {\em Boolean matrix theory and its applications}.
\newblock Marcel Dekker, 1982.

\bibitem{KR01}
Daniel Kobler and Udi Rotics.
\newblock Polynomial algorithms for partitioning problems on graphs with fixed
  clique-width (extended abstract).
\newblock In {\em SODA}, pages 468--476, 2001.

\bibitem{KMM95}
Jan Kratochv\'{\i}l, Paul~D. Manuel, and Mirka Miller.
\newblock Generalized domination in chordal graphs.
\newblock {\em Nord. J. Comput.}, 2(1):41--50, 1995.

\bibitem{perm3}
Ross~M. Mcconnell and Jeremy~P. Spinrad.
\newblock Modular decomposition and transitive orientation.
\newblock {\em Discrete Math}, 201:189--241, 1995.

\bibitem{MDSchordbip}
Haiko M\"uller and Andreas Brandst\"adt.
\newblock The np-completeness of steiner tree and dominating set for chordal
  bipartite graphs.
\newblock {\em Theor. Comput. Sci.}, 53:257--265, 1987.

\bibitem{thesisOum}
Sang\mbox{-}il Oum.
\newblock {\em Graphs of Bounded Rank-width}.
\newblock PhD thesis, Princeton University, 2005.

\bibitem{OS06}
Sang\mbox{-}il Oum and Paul~D. Seymour.
\newblock Approximating clique-width and branch-width.
\newblock {\em J. Comb. Theory, Ser. B}, 96(4):514--528, 2006.

\bibitem{GM10}
Neil Robertson and Paul~D. Seymour.
\newblock Graph minors. x. obstructions to tree-decomposition.
\newblock {\em J. Comb. Theory, Ser. B}, 52(2):153--190, 1991.

\bibitem{circperm1}
Doron Rotem and Jorge Urrutia.
\newblock {Circular permutation graphs.}
\newblock {\em Networks}, 12:429--437, 1982.

\bibitem{S03}
Jeremy Spinrad.
\newblock {\em Efficient Graph Representations}.
\newblock American Mathematical Society, Fields Institute, 2003.

\bibitem{circperm2}
R.~Sritharan.
\newblock A linear time algorithm to recognize circular permutation graphs.
\newblock {\em Networks}, 27(3):171--174, 1996.

\bibitem{TP97}
Jan~Arne Telle and Andrzej Proskurowski.
\newblock Algorithms for vertex partitioning problems on partial {\it k}-trees.
\newblock {\em SIAM J. Discrete Math.}, 10(4):529--550, 1997.

\bibitem{convex1}
Alan Tucker.
\newblock A structure theorem for the consecutive 1's property.
\newblock {\em Journal of Combinatorial Theory, Series B}, 12(2):153 -- 162,
  1972.

\bibitem{Tucker80}
Alan~C. Tucker.
\newblock An efficient test for circular-arc graphs.
\newblock {\em SIAM J. Comput.}, 9(1):1--24, 1980.

\bibitem{Yannakakis82}
Mihalis Yannakakis.
\newblock The complexity of the partial order dimension problem.
\newblock {\em SIAM Journal on Algebraic and Discrete Methods}, 3(3):351--358,
  1982.

\end{thebibliography}
%\end{scriptsize}

\appendix
\section{Appendix}

\subsection*{Interval graphs}

\begin{defi}
A graph is an interval graph if it has an intersection model consisting of
intervals on a straight line.
\end{defi}

Proof of Corollary \ref{cor_interval}.
\begin{pr}
Any interval graph has an interval representation where no interval starts or
ends at the same point. We build a caterpillar decomposition by sorting the
vertices by the left endpoint of their corresponding intervals. Let us now
consider a cut $(A,\overline{A})$ of the decomposition. Let $\sigma$ be the
total ordering of the vertices of $m(A)$ sorted by their right endpoint. 
Since all left endpoints of intervals corresponding to vertices of $A$ are to
the left of all left endpoints of intervals corresponding to vertices of
$\overline{A}$, two vertices $u \in A, u' \in \overline{A}$ are neighbours iff
the right endpoint of $u$ is to the right of the left endpoint of $u'$.
Hence $\forall u,v \in A :$ if $\sigma(u) \leq \sigma(v)$ then $N(u) \cap
\overline{A} \subseteq N(v) \cap \overline{A}$.
Then $\beta w(A) \leq \log(|m(A)|+1)$.
Since $m(A)$ is a clique in $G$ and there is one element in $\overline{A}$
neighbouring all vertices in $m(A)$, we have $|m(A)| \leq C-1$ hence the theorem
holds.
\end{pr}

\subsection*{Circular arc graphs}

\begin{defi}
A circular arc graph is the intersection graph of arcs of a circle.
\end{defi}

For more information on circular arc graphs see \cite{Tucker80}.

Proof of Corollary \ref{cor_circarc}
\begin{pr}
We define the starting (resp. ending) point of an arc as the first (resp. last)
point encountered when going around the circle in clockwise direction, starting
at a fixed point $p$.
We build a caterpillar decomposition by ordering the vertices according to the
startingpoint of their corresponding arc.
Now, we look at a cut $(A,\overline{A})$ of the decomposition. 
We show that the boolean-width of any circular arc graph is bounded by showing
that for any $S \subseteq A$ there is a set $S' \subseteq S$ of size at most $2$
such that $N(S) \cap \overline{A} = N(S') \cap \overline{A}$.
Assume $|S| > 2$, let $S'$ be the set containing the element of $S$ having the
first startingpoint and the element of $S$ having the last endpoint when
traversing clockwise starting from $p$.
Since no vertex of $\overline{A}$ can correspond to an arc properly contained in
the section between $p$ and the starting point of the element of $A$ with the
latest starting point, we have that $\forall u \in \overline{A}, u \in N(S)$ iff
$u \in N(S')$. 
Since $m(A)$ is the union of two cliques in $G$ and there exist two elements in
$\overline{A}$ neighbouring all vertices in $m(A)$, we have $|m(A)| \leq
2(C-1)$.
Thus, there can be at most $(2C)^2$ neighbourhoods and the theorem holds.
\end{pr}

\subsection*{Circular permutation graphs}

Proof of Theorem \ref{thm_circleprem}
\begin{pr}
We build a caterpillar decomposition using $\sigma_i$, an ordering obtained by
sorting the vertices by the inner endpoint of their corresponding line in
clockwise order starting with any point.
Let us now consider a cut $(A,\overline{A})$ of the decomposition.
Let $l$ be the unique line and $v$ the corresponding vertex such that:
\begin{enumerate}
\item $v \in A$
\item All lines corresponding to vertices appearing before $v$ in $\sigma_i$
cross $l$ in a clockwise direction.
\item No line corresponding to a vertex in $A$ appearing after $v$ in $\sigma_i$
cross $l$ in a counter-clockwise direction.
\end{enumerate}
Let $v$ be the first vertex of $\sigma_o$, an ordering of the vertices of
$m(A)$.
Continue the ordering by repeating the two steps above for the vertices in $A$
not yet in $\sigma_o$.

Since all inner endpoints of lines corresponding to vertices of $A$ are
consecutive on the inner cycle, if a vertex $u \in \overline{A}$ is neighbour
with a vertex $v \in A$ then either $u$ is neighbour with all vertices before
$v$ in $\sigma_o$ or $u$ is neighbour with all vertices after $v$ in $\sigma$.

If $S \subseteq A$ is a minimal set then $|S| \leq 2$. 
Assume for contradiction that there are at least $3$ elements in $S$.
Let $x,y,z$ be three elements of $S$ such that $\sigma_o(x) < \sigma_o(y) <
\sigma_o(z)$.
Now any vertex in $N(y) \cap \overline{A}$ will see either $x$ or $z$. Hence
$N(S \setminus y) = N(S)$ contradicting minimality of $S$.

The number of neighbourhoods is at most $n^2$, hence the theorem holds.
\end{pr}

\subsection*{Trapezoid graphs}

\begin{defi}
A graph is a trapezoid graph if it is the intersection graph of trapezoids
between two parallel lines.
\end{defi}

For more information on trapezoid graphs see \cite{trapez1}.

\begin{thm}
The boolean-width of a trapezoid graph $G$ is at most $2 \log n$.
\end{thm}

\begin{pr}
We build a caterpillar decomposition by sorting the vertices by the upper right
corner of their corresponding trapezoid from left to right.
Let us now consider a cut $(A,\overline{A})$ of the decomposition.
We show that the boolean-width of any trapezoid graph is bounded by showing
that any neighbourhood has a representative of size at most 2.
Let us take any $S \subseteq A$ with $|S| > 2$, we build a set $S' \subseteq
S$ with $N(S) \cap \overline{A} = N(S') \cap \overline{A}$ and $|S'|
\leq 2$: for the upper line (resp. lower), we take the the trapezoid $u$ (resp.
$l$) with the rightmost upper (resp. lower) right corner, we then set $S' =
\{u,l\}$.
Let us assume for contradiction that $\exists x \in \overline{A}: x \in N(S)
\backslash N(S')$. The trapezoid of $x$ must intersect some trapezoid of $S$ on
the upper or lower line. If it does not intersect $u$ or $l$, then the whole
trapzeoid of $x$ is to the right of $u$ and $l$.
By construction of the decomposition, $x$ would have been in $A$.
Thus $N(S) \cap \overline{A} = N(S') \cap \overline{A}$, and hence any
neighbourhood has a representative of size at most $k+1$ and the theorem
holds.
\end{pr}

\subsection*{k-trapezoid graphs}

Proof of Theorem \ref{thm_ktrapez}
\begin{pr}
We build a caterpillar decomposition by sorting the vertices by the rightmost
corner of their corresponding $k$-trapezoid. Let us now consider a cut
$(A,\overline{A})$ of the decomposition.
We show that the boolean-width of any $k$-trapezoid graph is bounded by showing
that any neighbourhood has a representative of size at most $k$.
Let us take any $S \subseteq A$ with $|S| > k$, we build a set $S' \subseteq
S$ with $N(S) \cap \overline{A} = N(S') \cap \overline{A}$ and $|S'|
\leq k$: for each line $i$, we take the the $k$-trapezoid $r_i$ such that
$r_i$ is the $k$-trapezoid corresponding to a vertex in $S$ containing the
rightmost point on the $i^{th}$ line. We set $S'$ as the set of all $r_i$ and
$l_i$.
Let us assume for contradiction that $\exists x \in \overline{A}: x \in N(S)
\backslash N(S')$. There must exist some $j$ such that the $k$-trapezoid
of $x$ intersects some trapezoid of $S$ on the $j^{th}$ line. If it does not
intersect $r_j$, then the whole trapezoid of $x$ is to the right of $r_j$.
By construction of the decomposition, $x$ would have been in $A$.
Thus $N(S) \cap \overline{A} = N(S') \cap \overline{A}$, and hence any
neighbourhood has a representative of size at most $k$ and the theorem
holds.
\end{pr}

\subsection*{Lower bounds}
Proof of Corrollary \ref{cor_HsuSC}
\begin{pr}
Let a Hsu-Stable-chain of length $q$ and width $p$ be the Hsu-join-chain of
length $q$ and width $p$ where $\forall i, G_i$ is a stable set of size $p$.
We now have to show that every Hsu-Stable-chain is a bipartite permutation
graph. Since it is trivial that every Hsu-Stable-chain is a bipartite graph
(we put the $G_i$ in one colour class when $i$ is even, and in the other colour
class when $i$ is odd), we just have to build the permutation representation of
any Hsu-Stable-chain.
For any $i \in \{1,...,q\}$, we represent $G_i$ as a set of $p$ parallel lines,
thus building a stable set of size $p$, the sets of lines going alternatively
top-down and bottom up, like shown on figure \ref{rank-bipperm}. We then make
the lines cross by extending the scheme shown on figure \ref{rank-bipperm}: for
any $G_i$ the $j^{th}$ line is crossed by the $j$ last lines of the
$(j-1)^{th}$ set and the $j$ first lines of the $(j+1)^{th}$ set, hence
building a Hsu-Stable-chain.
\end{pr}

Proof of Corrollary \ref{cor_HsuCC}
\begin{pr}
Let a Hsu-Clique-chain of length $q$ and width $p$ be the Hsu-join-chain of
length $q$ and width $p$ where $\forall i, G_i$ is a $K_p$.
We now have to show that every Hsu-Clique-chain is a unit interval graph, which
we do by building the unit interval representation of a $p \times q$
Hsu-Clique-chain. We build each $K_p$ of the Hsu-Clique-chain using $p$
intervals, each one being slightly shifted to the right w.r.t. the previous one,
like shown on figure \ref{rank-unitint}. We then put each set of intervals just
next to the previous one, without having the intervals at the same height
overlapping: for any $G_i$ the $j^{th}$ line is crossed by the $j$ last lines of
the $(j-1)^{th}$ set and the $j$ first lines of the $(j+1)^{th}$ set, hence
building a Hsu-Clique-chain.
\end{pr}

\end{document}